\documentclass[11pt]{article}
\usepackage{amsmath,amssymb,amsthm,mathtools}
\usepackage[margin=1in]{geometry}
\usepackage[colorlinks=true,linkcolor=blue,citecolor=blue]{hyperref}
\usepackage{pgfplots}
\pgfplotsset{compat=1.17}

\usepackage[
backend=biber,
style=nature,
]{biblatex}
\newcommand{\E}{\mathbb{E}}

\DeclareMathOperator{\sign}{sign}
\newcommand{\I}{\mathcal{I}}

\title{On Bonart's interpretation of the Square-Root Impact Law}
\author{
J.-P.~Bouchaud, I. Mastromatteo, B. T\'oth \\ 
{Capital Fund Management\footnote{23 rue de l'Université, 75007 Paris, France.}}}
\date{\today}

\begin{document}
\maketitle

\begin{abstract}
The square-root impact law (SRIL), $\I\simeq Y\sigma\sqrt{Q/V}$, bundles two facts
that a single mechanism must explain at once: a \emph{shape} (impact proportional to square-root of traded volume $Q$) and an \emph{amplitude} ($Y=O(1)$, independent of the participation rate
$\varphi$). Fixed-kernel models reproduce the shape but carry a ``$\sqrt\varphi$
wart,'' $Y \propto \sqrt{\varphi}$. Bonart has recently proposed an elegant escape \cite{bonart2026diffusive}: if realized
and counterfactual prices are both diffusive, information-neutral impact must have
white increments and the SRIL follows without the wart. We reformulate and simplify
his argument, and foreground the ingredient he himself regards as essential --- that
the market whitens the stationary \emph{stream} of each participant's metaorders,
not any isolated one. A solvable linear model (Lillo--Mike--Farmer) satisfies his
premises, yet gives an \emph{isolated} metaorder the super-square-root impact
$\varphi^{(1-\gamma)/2}Q^{(1+\gamma)/2}$, $\gamma\in(0,1)$, because  single metaorders (i.e. not part of a stream)
are statistically invisible and thus priced mechanically. We then construct an explicit
multi-agent propagator that realizes Bonart's whitening, and show it reproduces the
clean SRIL {precisely when} the market can attribute trades to their issuer. But correct re-attribution of anonymous trades is a formidable task, which requires markets to behave as implausibly efficient signal processing machines. The stream hypothesis makes a sharp, falsifiable prediction --- isolated metaorders, well separated in
time, should not obey the SRIL --- that does not seem agree with empirical data known to us.
Independently, a volume-conservation argument singles out the square-root law and lays bare the step the diffusivity route must assume: diffusivity fixes the \emph{variance} of impact --- an additive, stationary, age-blind quantity --- whereas the SRIL is a law for the \emph{mean} impact, which marginally decays as $1/\sqrt{\text{age}}$ and therefore demands an origin of time. Finally, the pure
all-pass phase left free by diffusivity pins the magnitude of permanent impact to
that of the peak, instead of the observed decay toward zero. Together, these point
away from whitening and toward a mechanical resolution: the square root reflects
\emph{volumes that must balance}, not \emph{correlations that must vanish}.
\end{abstract}

\tableofcontents

\section{Introduction}
\label{sec:intro}
The square-root impact law (SRIL hereafter)
\begin{equation}
\I \;=\; Y\,\sigma\,\sqrt{\frac{Q}{V}}
\label{eq:sril}
\end{equation}
relates the price impact $\I$ of a metaorder of size $Q$ to the daily traded
volume $V$ and the daily volatility $\sigma$, with a dimensionless prefactor $Y=O(1)$ \cite{bouchaud2018trades, webster2023handbook}.
Equation~\eqref{eq:sril} is among the most robust empirical regularities in all
of market microstructure: it holds across asset classes, venues, and epochs, and
it is remarkably insensitive to the details of how the metaorder is sliced and
scheduled, and to the type of trading strategy used~\cite{sato2024does} --- see the extended discussion in \cite{bouchaud2018trades} and refs.~therein. This robustness is itself a clue: a law that survives so much
heterogeneity is unlikely to depend on the fine microstructural mechanics that
differ from venue to venue. It must reflect something deeply structural.

However, the theoretical interpretation of the SRIL is still much open to debate.
The latent liquidity model (dubbed LLOB) proposed
in~\cite{toth2011anomalous,donier2015fullyconsistentminimalmodel,donier2016walras} offers a plausible framework that
naturally leads to Equation~\eqref{eq:sril}, but only in a high-participation
trading regime that does not seem to correspond to the empirical regime where the
SRIL is observed, typically $Q/V\lesssim0.1$. This discrepancy, and possible ways to escape
it, were addressed in Refs.~\cite{benzaquen2017marketimpactmultitimescaleliquidity,bucci2018slow,benzaquen2018fractional}, but nagging doubts on the
validity of their escape were expressed in~\cite{maitrier2025double}, based on the analysis of all transactions on the Tokyo Stock Exchange. We shall return to
the latent-book picture, and to whether it can be reconciled with the arguments of
the present note, in a companion paper; here we set it aside.

More recently, Bonart~\cite{bonart2026diffusive} proposed another, trivial-looking origin for
the SRIL, rooted in the diffusive nature of prices. His argument, which we review
and discuss at length below, is based on the idea that the impact of an
individual participant's flow is ``whitened'' by the market, so that prices
remain diffusive \emph{with} and \emph{without} that participant's metaorders.
With one extra assumption, Bonart derives Eq.~\eqref{eq:sril} exactly, and
correspondingly calls it the ``inconspicuous law of market impact.'' The purpose
of this note is to examine this proposal on its own terms: to isolate what it
genuinely secures, and what it assumes, either explicitly or tacitly. We shall argue that the missing
ingredient is \emph{detection} and, more pointedly, that the information processing
an anonymous market would need from the order flow it sees is so formidable that it
seems quite implausible, at least to our eyes.

It is important to note that \eqref{eq:sril} bundles together \emph{two}
structural facts that must be explained by the \emph{same} mechanism. This
imposes strong constraints on eligible theories, disqualifying those that account
for only one of them.

\paragraph*{Feature (i): the shape.} Impact grows as $\sqrt{Q}$. This is the
concave, ``diminishing marginal impact'' law: the tenth million shares move the
price $\sqrt{10} \approx 3.2$ less than the first million. It suggests that liquidity is not a
fixed depth to be consumed once, but a renewable resource that the very act of
trading calls forth (as assumed in the LLOB picture \cite{bouchaud2018trades}).

\paragraph*{Feature (ii): the amplitude.} The prefactor $Y$ is $O(1)$ and, most
tellingly, \emph{does not scale with the participation rate}
\begin{equation}
\varphi \;\equiv\; \frac{Q}{f T},
\label{eq:phidef}
\end{equation}
i.e. the rate at which the metaorder consumes volume (where $T$ is the time needed for fully executing the metaorder, and $f$ the volume traded by the market per unit time). Concretely: take two metaorders of the \emph{same} total size $Q$, one
executed patiently at $\varphi=0.1\%$ of market volume, the other more aggressively at
$\varphi=1\%$. Eq. \eqref{eq:sril} says that they incur essentially the \emph{same} impact. Said differently, the SRIL says that impact is only dependent on the traded volume $V$, not on the time $T$ it took for being executed. It is this
participation-independence --- on top of the square root per se --- that is the deep and
counter-intuitive fact.

Now, in the regime of small participation $\varphi$ the LLOB model boils down to a
simple linear propagator model \cite{donier2015fullyconsistentminimalmodel}, in which feature~(i) is reproduced but feature~(ii)
comes out wrong:
\begin{equation}
\I \;\sim\; \sqrt{\varphi\,Q}\,,
\label{eq:wart}
\end{equation}
i.e.\ $Y\propto\sqrt\varphi$. We call \eqref{eq:wart} the \emph{$\sqrt\varphi$
wart}. Another way to write Eq.~\eqref{eq:wart} is in
terms of $T$. Noting that
$Q \propto \varphi T$, we obtain $\I \;\sim\; Q/\sqrt{T}$. For the two metaorders above,
for which $T$ varies by a factor $10$ at fixed size $Q$, this predicts a factor
$\sqrt{10} \approx 3.2$ difference in impact --- a gross effect that would be
impossible to miss, and that is simply not there~\cite{bucci_impact_vol2019,maitrier2025double}. In fact
the wart is not a defect of one particular model but a near-theorem: for any
\emph{linear}, self-similar kernel $G(\tau)\sim\tau^{-\beta}$ the exponents of
$\varphi$ and $Q$ are locked by dimensional homogeneity to sum to unity,
$\I\sim\varphi^{\beta}Q^{1-\beta}$, so demanding the observed $\sqrt Q$
($\beta=\tfrac12$) forces the $\sqrt\varphi$. Escaping the wart therefore requires
leaving the linear, self-similar family altogether.

Bonart's proposal elegantly avoids the wart and correctly predicts a
$\varphi$-independent prefactor, by replacing the fixed mechanical kernel with the
requirement that impact accumulate \emph{diffusively}. In this note we first
restate Bonart's additive variance law without flourishes, and
the mean-equals-standard-deviation step through which the square root emerges free
of the wart (Sec.~\ref{sec:diffarg}); we also flag there the detectability problem
this step raises. 

We then confront the argument with a solvable linear model
(Sec.~\ref{sec:lmf}), which satisfies most premises of Bonart's neutral-regime
argument, yet fails to reproduce the SRIL for {\it individual} metaorders. This result highlights what Bonart emphasizes as a crucial ingredient in his construction: that the market is supposed to detect the statistical signature of a stationary {\it stream} of metaorders, not isolated ones. It is only this stream that the market is able to whiten, not any particular chunk. 

Section~\ref{sec:propagator} then asks whether a phenomenological
``perceived-persistence'' propagator could supply that detection. 
We construct an explicit, multi-agent propagator model that generates exactly Bonart's whitening device, but by the same token illustrates how formidable this assumption is. The detection hypothesis is also at the heart of three proposals made in the literature to explain the SRIL of {\it isolated} metaorders (Farmer--Gerig et al. \cite{farmer2013efficiency}, Saddier--Marsili \cite{saddier2024bayesian}, Maitrier--Bouchaud \cite{maitrier2025subtle}). The assumption that a somewhat regular stream of metaorders can be detected and attributed to a single market participant would then be a way to justify that single metaorders can be statistically detected on the fly. Section~\ref{sec:allpass} then shows that the decay of impact tells the same story from the phase side: diffusivity fixes only the spectral magnitude of the response, but its pure all-pass phase is empirically wrong, and the same perceived correlator would cure it --- at the same cost in detection.

Finally, Section~\ref{sec:volume} reframes the diffusivity argument in terms of a purely macroscopic volume-conservation constraint that singles out, but does not pin down, the SRIL. In particular, it makes a clear distinction between the {\it average} impact and its standard deviation, which are conflated in Bonart's argument. 
The result is suggestive: impact appears to be set by the \emph{volumes} a metaorder must transact, not by the
\emph{correlations} it induces. (As noted above, the SRIL only depends on $V$ and
not on $T$.) In Sec.~\ref{sec:remarks}, two short remarks show that mere diffusivity selects no particular impact law, even when detection is granted. 

\bigskip
\noindent\textbf{Notation.} We reserve $\varphi$ for the participation rate.
Bonart writes $\mu$ for the same quantity, and the literature sometimes uses $\nu$
for a cancellation rate; the LLOB reaction--diffusion picture also uses $\nu$ for
cancellation. To avoid future collisions we use $\varphi$ throughout for
participation, as in Ref.~\cite{maitrier2025subtle}. Throughout the main development we take
unit-volume child orders, i.e.\ $q=1$, so that ``number of trades'' $n$ and
``executed volume'' are often interchangeable. Most of the time, we also interchange real time $T$ and trade time $N$. We use $M$ to denote the total number of metaorders. 

\section{The diffusivity argument}
\label{sec:diffarg}

\subsection{The counterfactual decomposition and structural diffusivity}
\label{sec:diffusivity}
Bonart's starting point is a
\emph{counterfactual} decomposition of \emph{returns}. Let $x_t$ be the 
return over a bin of size $\Delta t$, and let $y_t$ be the \emph{counterfactual}
return that would have prevailed had the investor not traded. The investor's
market impact is the difference,
\begin{equation}
x_t \;=\; y_t + j_t,\qquad j_t \equiv x_t-y_t,
\label{eq:decomp}
\end{equation}
so that $y_t$ is the ``market-without-me'' and $j_t$ is the investor's own causal
footprint. This is a decomposition into \emph{background flow} and \emph{own
impact}: $y_t$ is not an ``efficient''
or ``fundamental'' price, it is the entire rest of the market in the investor's
absence.

Bonart's assumption is that \emph{both} the realized and the
counterfactual returns are diffusive. This is not imposed by fiat but
\emph{adaptively}: in the absence of persistent arbitrage, returns must be
serially uncorrelated, and arbitrageurs continuously recalibrate to the
statistical composition of order flow, restoring diffusivity as participants enter or exit. The counterfactual price --- the market without the investor --- is therefore also diffusive. 

Here we depart from Bonart's presentation and cut to the chase to offer a much simplified account of his construction that retains the gist of it.


Extending the argument above to all $M$ investors, we are led to conclude that the total return $x_t$ can be written as 
\begin{equation}
x_t \;=\; \xi_t + \sum_{k=1}^M j_t^{(k)},
\label{eq:decomp-total}
\end{equation}
where $j_t^{(k)}$ is the impact of the $k^{\rm th}$ investor trading at time $t$ and $\xi_t$ is any ``fundamental'' (i.e. not impact driven) component of price changes. If all these flows lead to a diffusive contribution, and provided flows are information neutral (i.e. uncorrelated with $\xi_t$ and with each other) then we immediately derive that the variance of price changes between $t=1$ and $t=N$ is given by
\begin{equation}\label{eq:variance}
\Sigma_N^2 = N \sigma_\xi^2 + b^2 s^2 \sum_{t=1}^N \sum_{k=1}^M \mathbb{I}(j_t^{(k)} \neq 0) = N \sigma_\xi^2 + b^2 s^2 \sum_{k=1}^M n_k,
\end{equation}
where $n_k$ is the number of orders executed by investor $k$, $b=O(1)$ and $b^2 s^2$ is the individual contribution of each trade to volatility, that we know is of the order of the bid-ask spread squared, $s^2$. Since $\sum_{k=1}^M n_k= N$ the total number of trades, we conclude that (i) $\Sigma_N^2  = N \sigma^2$ with $\sigma^2 = \sigma_\xi^2 + b^2 s^2$ and (ii) the contribution of each investor to volatility is proportional to their participation rate $n_k/N$. This is essentially the content of Bonart's equation (4.2) in Ref. \cite{bonart2026diffusive}. Note that since total volatility is found to be proportional to spread \cite{wyart2008relation, bouchaud2018trades}, one must also have $\sigma_\xi^2 \propto s^2$.

\subsection{Bonart's square root}
\label{sec:bonart-sqrt}

Now comes, in our view, Bonart's crucial assumption that \emph{the average impact of the metaorder is of the same order of magnitude as the standard deviation} computed above. This is far from a
trivial statement --- indeed, the case of a standard zero-drift random walk shows that such a conflation cannot hold in general; Sec.~\ref{sec:lmf} exhibits a solvable impact model in which the two differ completely, and Sec.~\ref{sec:volume} elaborates the point further.

Granting it for the moment, we get that the impact of metaorder $k$ when the market executes $N$ trades is given by:
\begin{equation} \label{eq:simple}
    I(n_k,N) \propto \varepsilon_k \sqrt{b^2 s^2 n_k},
\end{equation}
where $\varepsilon_k$ is the sign of the metaorder. Nothing in Bonart's construction fixes this sign and this is actually the crux of the problem, but let us accept fixing it to the direction of the metaorder. 

Now, in our simplified setting where all trades are of the same size $q=1$, one has $|Q_k| = q n_k = n_k$ and $V = q N = N$. Now, since $\Sigma_N^2 \propto N s^2$, one can rewrite Eq. \eqref{eq:simple} as $I(n_k,N) \propto \Sigma_N \sqrt{|Q_k|/V}$ or, rewriting in real time, the standard SRIL laid bare:
\begin{equation}
\I\big|_{\rm metaorder}\;\propto\;\sign(Q)\,\sigma \sqrt{\frac{|Q|}{V}}\,,
\label{eq:sril-bonart}
\end{equation}
independently of $\varphi$, i.e. without the ``wart'': the
$\sqrt\varphi$ that plagues the fixed-kernel models is here absent because the
participation rate $n_k/N$ and the duration combine into the total volume,
$\varphi\,T\propto Q$, so that $\varphi$ enters only through $Q$. The elegance is undeniable. The rest of this note asks what it entails.

Note that Bonart's argument also immediately explains why the impact profile of a metaorder is also a square-root, since one only has to replace $n_k$ in Eq.~\eqref{eq:simple} by $n_k^{(t)}$, i.e. the number of orders executed up to $t$, with $n_k^{(t=N)}=n_k$. Hence, without any additional assumption, both the peak impact and the running impact obey the square-root law, as empirically observed \cite{bouchaud2022inelastic, maitrier2025double}.

\subsection{Variance is stationary, but mean needs a start time}
\label{sec:detectability}

One should now make an important remark, that has to do with stationarity and when it breaks. Clearly, the variance law is stationary, in the sense that the extra contribution to volatility of each trade of any investor does \emph{not} depend on the time since the metaorder started. This is a direct consequence of the linearity of  Eq. \eqref{eq:variance}. But as soon as one takes the square-root of the variance to estimate the mean impact of each metaorder, one encounters a thorny problem that is intrinsic to the SRIL: the marginal impact of an additional trade \emph{must depend on the number of trades already executed since the start of the metaorder}, as $[{n_k^{(t)}}]^{-\frac12}$. In a nutshell: variance has no clock, while the mean requires one. 

Hence, in order to justify Bonart's identification of the mean impact and its standard deviation, one has to come up with a mechanism that allows the market to identify the start of a metaorder, at least to some level of precision. As we will discuss below, this is a problem shared with many of the previously proposed theories of the SRIL (except the LLOB model). 

\subsection{A note on interpretation}
\label{sec:interpretation}

Continuing on this detectability issue, Bonart forcefully insists on the fact his argument does not concern the impact of a single, isolated metaorder. Rather, each metaorder is viewed as a ``typical signed
excursion of stationary activity,'' and $\sqrt Q$ is the \emph{standard deviation}
of an accumulation of white increments, the mean and the standard deviation being explicitly conflated. If participants continually make round
turns in their inventory --- an Ornstein--Uhlenbeck-like churn of positions opened
and closed --- then a buy program is one excursion of an ever-present,
sign-balanced-on-average flow. 

Bonart's occasional language of a ``Brownian excursion'' is then meaningful: the stream is genuinely stationary, a metaorder is not a perturbation of it
but a typical fluctuation within it, and the excess correlation is a property of an ensemble the market sees every day and has purportedly already
estimated. In this limit the start time of a metaorder can be predicted, and the passage from variance $Q$ to mean $\varepsilon \sqrt Q$ is perhaps not too problematic. Indeed, the average height of a Brownian excursion of length $T$ is $\sqrt{T}$ (this is actually the core of the argument of Capponi \& Cont \cite{capponi2019trade}, see \cite{bucci_impact_vol2019} and footnote \ref{footnote:2}).

However, a metaorder that buys for a few days, does nothing for several weeks, and then sells for a few days --- the CFM-style execution, common to many investment strategies, for which metaorders are separated by times much longer than their execution time --- is not a stationary stream and cannot be read as an
excursion of an Ornstein-Uhlenbeck-like process: on the scale of its execution it is a rare, isolated coherent event
against an otherwise balanced (but long range correlated) background, and its onset is an actual event to be
located, not a recurring phase. For such a metaorder the start time cannot be pre-estimated as an ensemble property but something that would have to be inferred on the fly from a single realisation. 

It is worth separating two readings of this ``stream'' hypothesis, because they make
{different} predictions for an isolated metaorder. On the \emph{excursion}
reading --- continual round-turn (Ornstein--Uhlenbeck) churn, within which a
metaorder is a typical fluctuation --- an order that is well separated in time from
its predecessors is not an excursion of anything stationary, and should \emph{not}
obey the SRIL, as explicitly shown in the next section \ref{sec:lmf}. On the \emph{attribution} reading --- the market has learned to assign
trades to their issuer --- a recognizable repeat participant is whitened whenever
identified, \emph{regardless} of temporal separation, and Bonart's SRIL may survive. This reading thus yields a clear, testable prediction: \emph{A single metaorder starting long after the
previous one ended should not obey the SRIL under the excursion reading} --- a
smoking-gun prediction that does not seem to agree with CFM's empirical data, but we need to revisit this matter in the future  precisely because of its sharpness.

\section{Isolated metaorders: a linear counterexample}
\label{sec:lmf}
Bonart's result may read as though the SRIL followed from diffusivity alone. This section exhibits a fully solvable linear model in which the aggregate price is diffusive,
the amplitude law holds, and trades are information neutral --- yet the impact of a single metaorder is
\emph{not} the square root. Such a framework makes apparent that an extra ingredient is needed for Bonart's argument to make sense, which he argues is the statistical detectability of a \emph{stream} of metaorders from the same issuer. We will come back to this question in the next section \ref{sec:propagator}.

\subsection{A long-range correlated order flow}
Consider the Lillo--Mike--Farmer (LMF) mechanism~\cite{lillo2005theory}, in which the
long-range correlation of trade signs,
$C(\ell)=\E[\varepsilon_t\varepsilon_{t+\ell}]\sim\ell^{-\gamma}$
($\gamma\in(0,1)$), is generated by a broad, power-law distribution of metaorder
run-lengths: the metaorders \emph{are} the constituents of the flow. Let the
market price this flow through a \emph{linear}, participant-blind propagator $G$,
calibrated so that the realized price is diffusive:
$$
x_t
=
\xi_t + \sum_{t'= -\infty}^t G(t-t') \varepsilon_{t'}
=
\xi_t + \sum_{t'= -\infty}^t G(t-t') \sum_{k=1}^M \mathbb{I}(\varepsilon_{t'}^{(k)} \neq 0) \varepsilon^{(k)}_{t'} \,.
$$
For a linear model the return
autocorrelation is a functional of $C$ alone, and the whitening condition
$\hat g(\omega)\propto\hat C(\omega)^{-1/2}$ (with $g=\Delta G$) fixes the kernel
to the critical propagator\footnote{A pure power law for $G$ whitens the long-memory
\emph{tail} (Hurst $\tfrac12$) but leaves small residual autocorrelations at short
lags; exact whiteness is obtained with the associated causal all-pass filter,
which shares the low-frequency $\ell^{-(1-\gamma)/2}$ behaviour and hence leaves
the conditional-mean exponent below unchanged. The residual short-lag
non-whiteness of the bare power law is confined below the bid-ask spread scale and is not 
economically relevant, and could easily be cured by tweaking the short lag behaviour of $G$, see~\cite{lodhia2026impact}.}
\begin{equation}
G(\ell)\sim \ell^{-\beta},\qquad \beta=\frac{1-\gamma}{2}.
\label{eq:critical}
\end{equation}
In this market the price is diffusive; the excess variance carried by any
sub-flow of participation $\varphi$ is $\propto\varphi$, so the additive variance law
holds; and distinct participants are statistically independent, as required by Bonart.
Moreover the propagator whitens the participant's flow at the level of the stationary strategy's second-order statistics: the aggregate price is diffusive, so every premise Bonart places on the background is satisfied. What his argument additionally requires --- that a \emph{single coherent metaorder}, not the sign-alternating strategy, be whitened, incurring $\mathrm{Std}[\sum_t j_t^{(k)}]\sim\sqrt n_k$ --- is exactly the step we now show fails, even when the LMF model is a perfectly reasonable description of reality, as Sato et al. have recently shown \cite{sato2023inferring}.

\subsection{The undetected metaorder incurs under-concave impact}
Now inject \emph{one additional} metaorder, drawn from the same run-length
distribution, and recompute the price with the very same propagator. Because the
metaorders are the very constituents of the flow, adding one more is simply drawing
one more sample from the generating distribution: it changes no statistics, alters
$\gamma$ not at all, breaks no diffusivity, and creates no return predictability.
\emph{The metaorder is statistically invisible}, and the participant-blind
propagator prices it exactly like any other flow.

Its impact is then a deterministic mechanical convolution. A scheduled metaorder is
a \emph{coherent block} of $n=Q/q$ same-sign children, so the background flow ---
and all of its fluctuations --- cancels exactly in the counterfactual difference
$I_t = x_t-y_t$, leaving
\begin{equation}
{
I_t \;=\; \sum_{t'=1}^t G(t-t') \mathbb{I}(\varepsilon_{t'}^{(k)} \neq 0)\;}\sim\;\varphi^{\,\beta}Q^{\,1-\beta}
\Big|_{\beta=(1-\gamma)/2}\;=\;\varphi^{(1-\gamma)/2}\,Q^{(1+\gamma)/2},
\label{eq:lmf-impact}
\end{equation}
the last scaling being the same homogeneity relation $\I\sim\varphi^{\beta}Q^{1-\beta}$
that produced the wart (Eq.~\eqref{eq:wart}), now evaluated at the diffusive value
$\beta=(1-\gamma)/2$ rather than $\beta=\tfrac12$. At the empirical
$\gamma\approx\tfrac12$ this is $\I\sim\varphi^{1/4}Q^{3/4}$: under-concave in $Q$,
and carrying a residual $\varphi^{1/4}$. \emph{Both} features of the SRIL fail at
once.

\subsection{The missing ingredient is detection}

The gap between Eq. \eqref{eq:lmf-impact} and the SRIL is precisely what \emph{detection} would have to
close. Bonart's $\sqrt Q$ is the impact the run would incur \emph{if} the market
detected it and reshaped its response --- steepening the mean-response kernel from
the diffusive $\beta=(1-\gamma)/2$ to $\beta=\tfrac12$, and supplying the amplitude
that removes the residual $\sqrt\varphi$ (the mechanism we examine in
Sec.~\ref{sec:propagator}). This is an adaptive response, and it presupposes the metaorder has been somehow noticed --- in Bonart's picture because it belongs to a statistically identifiable stream of similar metaorders. In the LMF market, this means that some additional correlation between the sign of metaorders of a given length would need to be added. A generalisation of the LMF model that would take such correlations into account would be valuable to clarify exactly how Bonart's scenario would apply in that case. We take a short cut in the next Sect. \ref{sec:propagator} and posit that the market is indeed able to assign all metaorders in the order flow to individual market participants.

We note in passing that the under-concave, participation-dependent regime of the LMF model is not an
isolated pathology: it reappears in Sec.~\ref{sec:volume} as a degenerate corner
($b\to1$) of a macroscopic variance budget, the limit in which coherent metaorder impact
carries a vanishing ($O(M^{-1})$) share of the daily variance. The SRIL occupies the
complementary corner of that same budget, in which the share is $O(1)$ --- and, as we
shall see there, imposing it \emph{derives} the very mean-equals-standard-deviation
step that the present section shows diffusivity alone cannot supply.

\section{An effective multi-agent propagator model}
\label{sec:propagator}

Section~\ref{sec:lmf} leaves a central demand on any mechanism
that would realize Bonart's scenario for the SRIL: it must
steepen the conditional-mean response to $\beta=\tfrac12$, to turn the over-convex $Q^{(1+\gamma)/2}$ back into $\sqrt Q$. 

In this section we propose an explicit framework where this would happen, by assuming that each market participant is identified by the market, and assigned a certain participation rate $\varphi_k$. We believe this is the content of Bonart's model, and it shows how strong the underlying assumption really is. In particular, the way the market is able to find out who does what (even statistically) in an anonymous market is, at this stage, beyond us. It assumes a rather formidable information processing device for which we see no empirical evidence.\footnote{We have in particular repeatedly looked for statistical signatures of potential market reactions when CFM executes metaorders, and found very little, see e.g. \cite{toth2017short}, and I. Mastromatteo, unpublished (2015).} 

We then recall how the three most developed completions of the diffusive picture in the literature --- the efficiency/fair-pricing construction of Farmer, Gerig et al. \cite{farmer2013efficiency}, the Bayesian construction of Saddier and Marsili~\cite{saddier2024bayesian} and the two-time propagator of Maitrier and Bouchaud~\cite{maitrier2025subtle} --- all assume that the start of individual metaorders can be somehow detected. 

\subsection{A tentative construction}
\label{sec:perceived-C}
Suppose the market whitens returns using a propagator $G^{(k)}$ adapted to each market participant. In other words, we generalize Eq. \eqref{eq:lmf-impact} and write:
\begin{equation}
{
x_t = \xi_t + \sum_{t'=1}^t \sum_{k=1}^M G^{(k)}(t-t')  \mathbb{I}(\varepsilon_{t'}^{(k)} \neq 0)\;\varepsilon_{t'}^{(k)}}
\label{eq:gen-propagator}
\end{equation}
The market participant propagator $G^{(k)}$ is tied to the
\emph{reconstructed} sign correlator $C^{(k)}(\ell)$ by the diffusive-price condition. The idea is that a persistent directional flow, once detected (i.e. after time $\ell^\ast_k$), transiently \emph{flattens}
the perceived correlator: over lags up to the metaorder's current age $t'$ --- the
(trade-)time elapsed since its first child --- the correlator is held at a plateau
$C^\ast_k$ before the ambient decay resumes beyond $t'$:
\begin{equation}
C^{(k)}(\ell)\;\simeq\;
\begin{cases}
\text{ambient } C_0\,\ell^{-\gamma}, & \ell\lesssim \ell^\ast_k,\\[2pt]
C^\ast_k, & \ell^\ast_k\lesssim \ell\lesssim t',\\[2pt]
\text{ambient again}, & \ell\gtrsim t'.
\end{cases}
\label{eq:perceived-C}
\end{equation}
At completion the plateau extends to the full duration, $t'\to Q/\varphi_k$. Let us stress that $(k)$ labels a \emph{statistically coherent} stream, not a named or legally identified participant: the requirement is that the anonymous tape be \emph{unmixed} into persistent per-trader components, each with its own participation $\varphi_k$ --- a blind source-separation problem on the order flow, nothing more and nothing less.

In the following we assume (as does Bonart) that different market participants trade independently, i.e. $\mathbb{E}[\varepsilon_{t'}^{(k')}\varepsilon_{t''}^{(k'')}] \propto \delta_{k'k''}$. Whitening then fixes the propagator $\hat g(\omega)\propto\sigma\,
\hat C(\omega)^{-1/2}$ with $g=\Delta G$~\cite{bouchaud2003fluctuations,bouchaud2009markets, jusselin2018noarbitrageimpliespowerlawmarket}. (We drop the superscript $(k)$ hereafter to lighten notations.) 

On the correlation plateau ($\gamma_{\rm eff}=0$) this induces the critical build-up exponent of the propagator
$\beta_{\rm eff}=\tfrac12$, while the ambient tail gives $\beta=(1-\gamma)/2$. The propagator
amplitude scales as $C^{\ast\,-1/2}$, so that on the plateau $G(\ell)\sim \sigma\,C^{\ast\,-1/2}\,\ell^{-1/2}$. Convolving $G(\ell)$ with the coherent flow (density $\varphi$ per trade) up to the current age $t'$ -- and using the fact that metaorder signs are uncorrelated -- gives the \emph{running} impact due to a single metaorder
\begin{equation}
\I(t')\;\sim\;\int_0^{t'}\varphi\,G(\ell)\,d\ell
\;\sim\; \varphi\,\frac{\sigma}{\sqrt{C^{\ast}}}\int_{0}^{t'}\ell^{-1/2}\,d\ell
\;\sim\;\sigma\,\sqrt{\frac{\varphi}{C^\ast}}\,\sqrt{\varphi\,t'},\qquad (t' \gtrsim \ell^\ast)
\label{eq:prop-running}
\end{equation}
a square-root \emph{profile} in the executed volume $\varphi\,t'$. At completion ($\varphi\,t'=Q$) this is the peak impact
\begin{equation}
\I\;\sim\;\sigma\,\sqrt{\frac{\varphi}{C^\ast}}\,\sqrt Q,
\label{eq:prop-wart}
\end{equation}
so that in SRIL form $Y\propto\sqrt{\varphi/C^\ast}$ and
\begin{equation}
\boxed{\;\text{wart-free, }\varphi\text{-independent SRIL}
\;\Longleftrightarrow\;C^\ast\propto\varphi\;.}
\label{eq:Cstar-phi}
\end{equation}
If the trading style of each participant, and in particular its trading intensity $\varphi_k$, can be identified, the relation $C^\ast\propto\varphi$ is indeed fully justified: conditioned to one child order being identified as belonging to a metaorder of type $(k)$, the probability that another child order belongs to the same metaorder is $\varphi_k$.

The tail beyond $t'$ --- the resumption of the ambient decay --- is what describes the post-execution relaxation
of impact and is a possible resolution of the paradox described in  Sec.~\ref{sec:allpass}. We now show that absent the mechanism invoked by Bonart, i.e. that metaorder regularities allow each of them to be identified, an anonymous market would deliver $C^\ast\propto\varphi^2$.

\subsection{The anonymous plateau is $\varphi^2$, but is unresolvable at small $\varphi$}
\label{sec:estimator}
If no one can work out whether a metaorder is present, the only object the market can respond to is a statistic of the
tape; since the sign is unknown the drift is unavailable, and the natural signal is
the excess sign correlation the plateau is meant to be. During the life span of a metaorder of participation
$\varphi$, the probability that two orders come from the metaorder is $\varphi^2$, so
it self-detects a flat excess
\begin{equation}
C^\ast\sim\varphi^2,
\label{eq:Cstar-honest}
\end{equation}
not the $\varphi$ that Eq. \eqref{eq:Cstar-phi} requires. Fed back, this gives
$Y\propto\varphi^{-1/2}$: an ``anti-wart'' in which impact \emph{decreases} with
participation.\footnote{\label{footnote:2} This anti-wart is equivalent to Capponi \& Cont's reading of the square-root law \cite{capponi2019trade}, in which impact is simply
identified with the realised absolute price move, $\mathcal I\sim\langle|\Delta
p|\rangle\sim\sigma_1\sqrt{N}=\sigma_1\sqrt{Q/\varphi}$, over the metaorder's
lifetime $N=Q/\varphi$; see the discussion in \cite{bucci_impact_vol2019}.} 

Worse, the plateau generally cannot be resolved at all, and least of all where the
SRIL is cleanest. Detection is on the flow: over a window of $n$ trades the
metaorder deposits a coherent net sign $\varphi n$, which must stand above the
fluctuations of the net sign of the ambient flow. With a correlated background
$C(\ell)=C_0\,\ell^{-\gamma}$ the latter has variance
\begin{equation}
V_n=\sum_{i,j\le n}C(i-j)=\underbrace{n}_{\text{uncorrelated count}}
+\underbrace{C_0\,n^{2-\gamma}}_{\text{correlated tail}},
\label{eq:Vn}
\end{equation}
and detection at signal-to-noise unity is the single balance
\begin{equation}
\boxed{\;(\varphi n)^2\;=\;n\;+\;C_0\,n^{2-\gamma}\;.}
\label{eq:balance}
\end{equation}
The two terms on the right are physically distinct noise floors, and which one
controls the crossover volume $Q^\ast=\varphi\,\ell^\ast$ depends on $\gamma$ and on
the amplitude $C_0$.

\emph{Diffusive regime} ($\gamma>1$, or $\gamma<1$ with $n$ below the correlation
scale). The count term dominates, $V_n\simeq n$; the correlations merely renormalise
the diffusion constant. Balancing $(\varphi n)^2=n$ gives
\begin{equation}
\ell^\ast\sim\varphi^{-2},\qquad Q^\ast\sim\varphi^{-1},
\label{eq:Qdetect-SR}
\end{equation}
which is $C_0$-independent to leading order and is the cheapest possible detection
volume. This is the regime in which the market is effectively diffusive; it is where
the square-root law is recovered without incident, and (in the short-range case) it
is the regime addressed by inference-based accounts such as that of Saddier and Marsili~\cite{saddier2024bayesian}.

\emph{Correlated regime} ($\gamma<1$, $n$ above the correlation scale). The tail wins,
and balancing $(\varphi n)^2=C_0\,n^{2-\gamma}$ gives
\begin{equation}
\ell^\ast\sim C_0^{1/\gamma}\varphi^{-2/\gamma},\qquad
Q^\ast\sim C_0^{1/\gamma}\,\varphi^{-(2-\gamma)/\gamma}
\;=\;C_0^{2}\,\varphi^{-3}\quad(\gamma=\tfrac12).
\label{eq:Qdetect-LR}
\end{equation}
Note that this detection scale is precisely the lower crossover $\ell^\ast$ of the
correlator \eqref{eq:perceived-C}: with $C^\ast\sim\varphi^2$, the ambient decay
meets the plateau at $C_0(\ell^\ast)^{-\gamma}=C^\ast$, i.e.\ $\ell^\ast\sim
C_0^{1/\gamma}\varphi^{-2/\gamma}$ --- the estimator can resolve the plateau only
beyond the very lag at which the plateau would begin. The two regimes meet where $C_0\,n^{1-\gamma}\sim1$, i.e.\ at
$n_\times\sim C_0^{-1/(1-\gamma)}$. This matters quantitatively: empirical sign
correlations are long-range but often with a \emph{small} amplitude, $C_0$ of order $10^{-2}$--$10^{-1}$, so $n_\times\sim C_0^{-2}$ (at $\gamma=\tfrac12$) can be large. A market
whose order-flow correlations have $C_0\simeq 0.01$ behaves diffusively for detection
purposes out to $n_\times\sim 10^{4}$ trades, so the cheap diffusive threshold
\eqref{eq:Qdetect-SR} governs a wide range of window sizes before the correlated
floor \eqref{eq:Qdetect-LR} takes over. The two are not alternatives but the two
ends of the single balance \eqref{eq:balance}.

Either way the conclusion is the same, and it sharpens as $\varphi\to0$, as  $Q^\ast$ \emph{grow unbounded}: for $\gamma= \frac12$, the coherent order is invisible to any
correlation-based estimator up to $\ell^\ast \sim 100$ in a favourable case $C_0=0.01$ and $\varphi=10 \%$ (i.e in the diffusive regime) but $\ell^\ast \sim 10^6$ for $C_0=0.1$ and $\varphi=1 \%$ (i.e in the correlated regime, with $n_\times\sim 10^{2}$), where the SRIL holds most cleanly.

The conclusion of this subsection is that without Bonart's ``stream'' hypothesis that makes individual metaorders potentially detectable, the metaorder-specific propagator assumption, Eq. \eqref{eq:perceived-C}, is untenable. This allows us to put other proposals based on metaorder identification in perspective.

\paragraph{A caveat: smart detection.} 
Eq. \eqref{eq:balance} sets the
threshold for a \emph{correlation-based} estimator working on signed flow alone. A
real market is not so impoverished: participation, volume and timestamp
regularities, order-size clustering, venue and broker footprints, and a host of
other HFT signals might carry information about a metaorder that the sign autocorrelation
discards. Any such channel improves detectability --- possibly substantially, and this is
presumably how execution-detection algorithms operate in practice. But the reduction is bounded by whatever coherence the metaorder actually
leaves on the tape, and a well-executed order is engineered precisely to leave as
little as possible --- whereas the SRIL appears universal and largely independent of the style of execution \cite{sato2024does}.

\subsection{Three realizations where detection re-enters}
\label{sec:realizations}
Three alternative completions of
the diffusive picture in the literature also reach the SRIL successfully, and it is
worth seeing precisely how. In all three, detection is not derived --- it is built into
the setup, through the same assumption we have found to be crucial and difficult to justify: that the
metaorder's onset (and hence its age) is known.

\paragraph{Farmer, Gerig, Lillo and Waelbroeck: efficiency plus fair pricing.}
The earliest of these completions~\cite{farmer2013efficiency} is also the closest in
spirit to our counterexample of Sec.~\ref{sec:lmf}, and pinpointing the difference
is instructive. Farmer et al.\ take the same long-memory order flow and impose two
conditions: market efficiency (returns are martingale) and a \emph{fair-pricing}
condition --- that the average price paid over the course of a metaorder equals the
permanent price it leaves behind, so that liquidity providers break even against it.
The efficiency condition alone is precisely what our LMF market embodies: it fixes
the participant-blind response that whitens the flow, and on its own it delivers the
over-convex mechanical impact of Sec.~\ref{sec:lmf}. The fair-pricing condition is
the extra ingredient, and it is what bends the impact toward the observed concave
law. But fair pricing is imposed order by order: it equates a metaorder's average
execution price with its own terminal impact, which presupposes that the pricing
``knows'' a metaorder is running, and over what horizon and size --- exactly the
conditioning on a coherent order that an anonymous tape does not supply. In our
language, Farmer et al.\ obtain the concave law by promoting detection to a
no-arbitrage \emph{axiom} (fair pricing) rather than deriving it: the metaorder's
presence and extent are built into the break-even condition from the outset. 

\paragraph{Saddier and Marsili: an explicit Bayesian detector.}
Saddier and Marsili~\cite{saddier2024bayesian} make the detection explicit. Their market
maker does not price through a mechanical kernel at all; it holds a
\emph{Bayesian posterior} over the presence and rate $\varphi$ of an informed trader,
and the price is the posterior mean of fundamental value given the observed flow.
Impact is then the \emph{revision} of that posterior as the order runs. Two
consequences matter directly for our argument. First, because a Bayesian estimate of a
persistent drift sharpens as $\propto1/\sqrt t$,\footnote{Until $t^{-1/2} \propto \varphi$, which coincides with the detectability threshold for short-range correlated flow, Eq. \eqref{eq:Qdetect-SR} above.} the
posterior value-revision inherits the square root: the concavity is the shrinking of
a belief, not the decay of a kernel, and the residual participation dependence is
absorbed into the volatility normalisation $\sigma\propto\sqrt\varphi$ (their Eq.~5) ---
Bonart's amplitude-law, inherited here from the assumption that informed traders adapt their participation rate to their signal. 

Crucially, their $\sqrt Q$
is a posterior \emph{mean}, not a root-mean-squared, so their construction is free of the
mean-versus-standard-deviation identification we flagged in
Sec.~\ref{sec:interpretation} as Bonart's weak point --- a genuine advantage.
The price of it is that detection is \emph{assumed}: their trader is informed and
sets $\varphi$ accordingly, and the onset of the metaorder ($t=0$) is taken to be identifiable.
The mechanism does not manufacture detection from an anonymous tape; it posits an
informed-trading equilibrium in which detection is automatic --- exactly the
ingredient we have argued is missing. And, tellingly, the construction inherits the
same fragility: fed long-range order flow, it too leaves the clean regime and
returns an under-concave impact (their Appendix~D.3). The exponent is different from ours but the qualitative lesson is identical: a detection-based square root is fragile
precisely when the background is realistically persistent.

\paragraph{Maitrier and Bouchaud: an age-clock propagator.}
The two-time propagator of~\cite{maitrier2025subtle} is the phenomenological realization of
exactly the object Sec.~\ref{sec:perceived-C} posited. Its kernel depends not only
on the lag but on the \emph{age} of the metaorder --- the time $t'$ elapsed since
its first child --- decaying more slowly at short age and reverting to the ambient
exponent at long age, and it robustly reproduces the $\sqrt Q$ peak with a
$\varphi$-independent prefactor. The age variable $t'$ is precisely an encoding of detection: to apply an
age-dependent kernel the market must somehow know that a metaorder started, and when. An
anonymous market reading the tape does not possess $t'$ --- it would have to infer
the onset, which is the very detection Sec.~\ref{sec:estimator} showed to be
unavailable below $Q\sim\varphi^{-3}$ (for $\gamma = \frac12$). The two-time propagator is therefore an effective description of a market that \emph{has} somehow detected the order and is
tracking its age. Read the other way,
its empirical success might be indirect evidence that something in real markets
does supply the onset --- that detection, by whatever route, has effectively
occurred --- which is precisely what Bonart's stream detectability aims at producing.

\section{The all-pass impact decay problem}
\label{sec:allpass}

Note that the detection gap has a counterpart on the decay side. Bonart's requirement that a
participant's own impact accumulate diffusively means its increment sequence is
white; writing the impact response to unit trade innovations as a causal filter
$U$, so that the increment spectrum is $|U(e^{i\omega})|^2$, whiteness reads \cite{bonart2026diffusive}
\begin{equation}
\big|U(e^{i\omega})\big|=1\qquad\forall\,\omega,
\label{eq:white}
\end{equation}
a pure (causal) all-pass condition. It constrains only the \emph{magnitude} of the
response, leaving its \emph{phase} undetermined. This is a genuine strength of the
framework --- ``white impact increments'' does not force ``impact equals
contemporaneous surprise,'' as in the propagator model \cite{bouchaud2003fluctuations}, and there is room for transient
build-up and relaxation. But the freedom is not harmless, and its pure form is inconsistent with empirically observed relaxation from peak to permanent impact.

\paragraph{An exact rigidity of pure all-pass filters.} The permanent
(long-horizon) impact of a unit innovation is the DC gain
$I_\infty=G_0 \sum_t U[t]=G_0\, U(e^{i0})$, where $U$ is a unit-energy all-pass,
$\sum_t U[t]^2=1$. Since $\omega=0$ is one of the frequencies at
which $|U|=1$,
\begin{equation}
\big|I_\infty\big|=G_0\,\big|U(e^{i0})\big|=G_0,
\qquad\text{while}\qquad
\big|I[0]\big|=G_0\,|U[0]|<G_0
\label{eq:rigidity}
\end{equation}
(the strict inequality because $\sum_t U[t]^2=1$ with more than one nonzero tap forces $|U[0]|<1$). The permanent
impact thus has magnitude pinned to $G_0$ --- i.e.\ to the very scale of the
peak --- while the instantaneous slippage is strictly smaller. What the
all-pass phase leaves free is only the \emph{sign} of $I_\infty$. 

The
first-order exponential family already exhibits both possibilities: with
$U(z)=(z^{-1}-a)/(1-az^{-1})$ with $|a| < 1$ so that the step response is $I(n)=G_0(1-(1+a)a^{n})$, hence an instantaneous impact $G_0 I(0) = - G_0 a$ that must be positive for consistency.
\begin{itemize}
    \item For $a \in ]-1,0]$, $G_0 > 0$ and impact climbs
\emph{monotonically from $I(0)= G_0 |a|$ to $I_\infty=G_0$}  --- so the impact reaches the end of
execution still short of its permanent level and then creeps \emph{up} to a plateau \emph{above} the peak.
\item For $a \in [0,1[$, one must now choose $G_0 < 0$ and the response reverts through zero to a plateau 
\emph{opposite} to the peak.
\end{itemize}  
In neither case ---  because $|I_\infty|=G_0$
throughout, nowhere in between --- can the permanent impact be made small compared
with the peak.

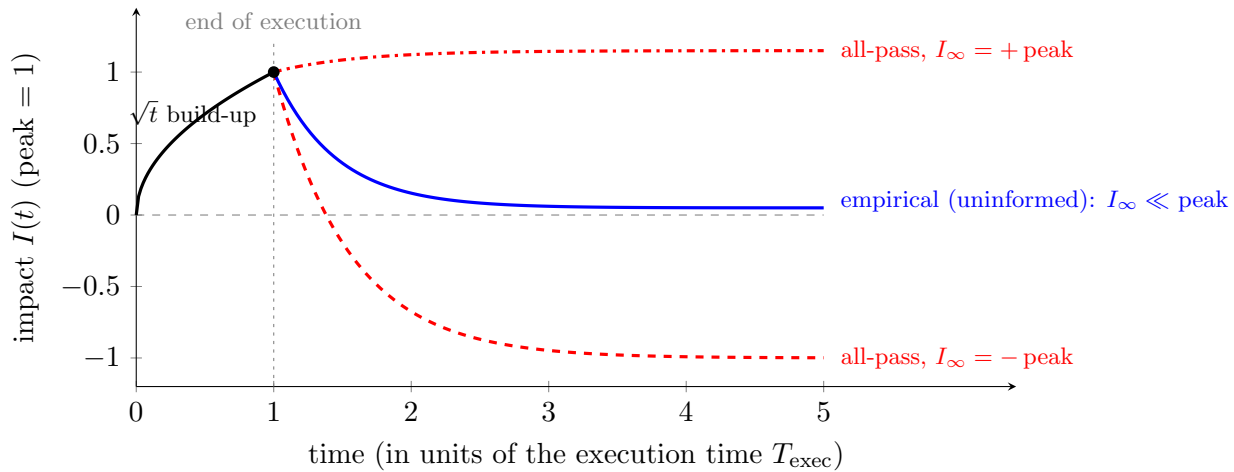
\begin{figure}[t]
\centering
\begin{tikzpicture}
\begin{axis}[
    width=0.80\textwidth, height=6.6cm,
    xlabel={time (in units of the execution time $T_{\rm exec}$)},
    ylabel={impact $I(t)$ (peak $=1$)},
    xmin=0, xmax=6.4, ymin=-1.2, ymax=1.45,
    axis lines=left,
    xtick={0,1,2,3,4,5},
    ytick={-1,-0.5,0,0.5,1},
    yticklabel style={/pgf/number format/fixed},
    every axis plot/.append style={very thick},
    clip=false,
]
\addplot[gray, dashed, thin] coordinates {(0,0) (5,0)};
\addplot[gray, dotted, thin] coordinates {(1,-1.2) (1,1.2)};
\node[gray, anchor=south, font=\footnotesize] at (axis cs:1.0,1.23) {end of execution};

\addplot[black, domain=0:1, samples=60] {sqrt(x)};

\addplot[red, dash dot, domain=1:5, samples=120] {1.15 - 0.15*exp(-(x-1)/0.6)};

\addplot[red, dashed, domain=1:5, samples=120] {-1 + 2*exp(-(x-1)/0.55)};

\addplot[blue, domain=1:5, samples=120] {0.05 + 0.95*exp(-(x-1)/0.45)};

\addplot[mark=*, mark size=1.6pt, only marks, black] coordinates {(1,1)};

\node[red,  anchor=west, font=\footnotesize] at (axis cs:5.05,1.15)
      {all-pass, $I_\infty=+\,$peak};
\node[blue, anchor=west, font=\footnotesize] at (axis cs:5.05,0.10)
      {empirical (uninformed): $I_\infty \ll$ peak};
\node[red,  anchor=west, font=\footnotesize] at (axis cs:5.05,-1.0)
      {all-pass, $I_\infty=-\,$peak};
\node[black, anchor=south east, font=\footnotesize] at (axis cs:0.95,0.55)
      {$\sqrt{t}$ build-up};
\end{axis}
\end{tikzpicture}
\caption{\small{Two poles of post-execution impact decay; the impact at the end of execution
(the peak of the build-up, dot) is normalised to $1$, and all curves share the concave
$\sqrt{t}$ build-up. They differ only afterwards. \emph{Empirically} (blue), impact
relaxes to a level of the \emph{same} sign as the peak but much smaller in magnitude
--- close to zero for uninformed flow \cite{gomes2015market, brokmann2015slow, bucci2018slow, bouchaud2022inelastic}. A \emph{pure causal all-pass} --- the generic
phase left undetermined by the diffusivity/white-increment
condition~\eqref{eq:white} --- has permanent impact of magnitude pinned to the peak,
$|I_\infty|=G_0=1$ (Eq.~\eqref{eq:rigidity}, with $|a|$ close to 1); only its \emph{sign} is free. The exponential
family thus contains a member whose step response climbs monotonically to $+1$ from
below, so that impact keeps drifting \emph{up} after execution to a plateau slightly
\emph{above} the peak (red dash-dotted), and a member that reverts through zero to a
plateau equal and \emph{opposite} to the peak (red dashed). The empirically observed
near-zero permanent level lies in the gap between these two poles, which no pure
all-pass can reach.}}
\label{fig:allpass}
\end{figure}

\paragraph{This is upside-down empirically.} Real metaorder impact builds up during
execution, \emph{peaks} near its end, and then relaxes to a permanent level that is
\emph{smaller in magnitude than, and of the same sign as, the peak} --- and, for
uninformed flow, close to \emph{zero}~\cite{gomes2015market,brokmann2015slow,bucci2018slow} (but see also \cite{benzaquen2017marketimpactmultitimescaleliquidity, bouchaud2022inelastic}). This decay toward a small fraction of the peak is precisely what a pure all-pass
cannot produce: its permanent impact is locked at full magnitude, whether at
$+\,$peak or $-\,$peak, and the empirically observed near-zero level sits in the gap
between these two poles that the all-pass phase can never reach
(Fig.~\ref{fig:allpass}). As Bonart himself notes, reproducing the decay ``calls for
structure beyond the flat-magnitude condition.''

The perceived correlator introduced in Sec.~\ref{sec:perceived-C} is in fact the
natural object that would cure this. After the end of the metaorder its tail
resumes, and if it resumes more steeply than the ambient value the impact relaxes to
a small permanent level of the same sign as the peak --- the blue profile of
Fig.~\ref{fig:allpass} rather than either red pole. The very construction that would
steepen the build-up to $\sqrt Q$ thus also shapes the decay correctly, from a single
object. But it inherits the very same liability: the plateau-and-tail must be keyed
to the metaorder's onset and extent, which --- as Sec.~\ref{sec:estimator} showed ---
an anonymous market cannot resolve in the low-participation regime where the SRIL
holds. The decay problem and the detection problem are, once again, one problem:
diffusivity fixes the magnitude of the response and leaves its phase free, and
pinning that phase to the empirically correct shape requires exactly the conditioning
on a coherent order that we have argued the market cannot supply.

\section{A volume argument that singles out the square-root}
\label{sec:volume}
We now present a purely \emph{macroscopic} argument --- using only conservation of
traded volume and the demand that the SRIL prefactor be universal and $O(1)$ ---
that singles out the square-root exponent independently of everything above, and
gives (in our eyes) the most convincing way to rephrase Bonart's argument relating
volatility and the SRIL, as it clarifies the mean-equals-standard-deviation identification that Sec.~\ref{sec:bonart-sqrt} had to
assume. This argument, however, does not mean that the SRIL is a foregone conclusion that follows from the diffusivity of prices alone. In particular, one recovers the LMF counterexample
of Sec.~\ref{sec:lmf} as a degenerate limit.

\paragraph{Setup.} Consider the metaorders executed on a trading day: 
$M$ of them,
with signed sizes $\varepsilon_k |Q_k|$ and number of child orders $n_k$, against total daily volume $V$, daily total number of trades $N$ and daily
volatility $\sigma$. Posit an end-of-day power-law impact
\begin{equation}
I_k=\varepsilon_k\,Y\,\sigma\left(\frac{|Q_k|}{V}\right)^{\delta} + b \, s \sqrt{n_k} \,\xi_k,
\label{eq:impact}
\end{equation}
where the first term is the average (signed) impact, and the second term the contribution of background volatility, with $s$ the bid-ask spread, $b \leq 1$ a certain coefficient, and $\xi_k$ IID random variables of mean zero and variance unity. Note that $Y$ captures, in an effective way, the impact decay from the end of execution to the end of day. 

We ask which value of $\delta$ is consistent with the requirement that the metaorders'
price moves account for the observed daily variance. To keep things as simple as possible, we assume that metaorders are mutually independent, with signs of mean zero, so that cross-metaorder terms vanish in the variance budget below.\footnote{This discards any contemporaneous correlation between distinct metaorders on the same name --- herding, or one metaorder acting as another's counterparty. Such correlations would reintroduce a cross term; we set them aside deliberately, not by oversight. In fact, as discussed in \cite{maitrier2025subtle}, such a correlation between metaorders is crucial to restore diffusivity when impact decay is accounted for.}

\paragraph{Variance budget and closure.} Writing the daily variance as accumulated
impact,
\begin{equation}
\sigma^2=\sum_k\langle I_k^2\rangle = Y^2 \sigma^2 \sum_k \Big(\frac{|Q_k|}{V}\Big)^{2\delta} + b^2 s^2 \sum_k n_k.
\end{equation}
Number and volume conservation give the exact identities
\begin{equation}
\sum_k n_k \equiv N; \qquad \sum_kQ_k \equiv V\quad\Longrightarrow\quad M\langle Q/V\rangle=1,
\label{eq:conservation}
\end{equation}
Noting that $N s^2 = \sigma^2$ \cite{wyart2008relation, bouchaud2018trades} and eliminating $\sigma$ we get
\begin{equation}
1 - b^2 = Y^2\,R(\delta),\qquad
R(\delta)\equiv
\frac{\big\langle(Q/V)^{2\delta}\big\rangle}{\big\langle Q/V\big\rangle}.
\label{eq:final}
\end{equation}
A better way to write this equation is to introduce a random variable $x = O(1)$ such that $|Q_k| = x_k V/M$, such that $\langle x\rangle =1$ (i.e. all metaorder volumes are on the scale of $V/M$) and
\begin{equation}
R(\delta)=
\frac{C_\delta}{M^{2 \delta - 1}}, \qquad C_\delta:=\big\langle x^{2\delta}\big\rangle = O(1),
\label{eq:final2}
\end{equation} 
such that 
\begin{equation}
 C_\delta Y^2 M^{1 - 2\delta}  = 1 - b^2 .
\label{eq:master}
\end{equation}
Equation~\eqref{eq:master} is the whole content of the argument, and the coefficient
$b$ --- the fraction of each trade's contribution to volatility that is
\emph{not} captured by the coherent mean impact --- selects between two worlds.

\begin{itemize}
    \item {\bf $b \to 1$: coherent impact is a subdominant $O(M^{-1})$ contribution to variance}. Here the mean impact carries a vanishing share of the daily variance, and \eqref{eq:master} forces $Y^2 M^{1-2\delta}\to 0$, i.e. $\delta>\tfrac12$ with a $\varphi$-dependent prefactor. This is \emph{exactly} the LMF corner of Sec.~\ref{sec:lmf}: setting $\delta-\tfrac12=\gamma/2$ and using $\varphi=1/M$ gives $Y\propto\varphi^{(1-\gamma)/2}=M^{\delta-1}$,
    whence $Y^2M^{1-2\delta}=M^{-1}\to0$, consistent with $b\to1$. The under-concave, participation-dependent impact $\varphi^{(1-\gamma)/2}Q^{(1+\gamma)/2}$ is thus not a separate phenomenon but the degenerate limit of the present budget in which coherent impact contributes negligibly to volatility. More generally, insisting on $Y=O(1)$ in this world would require $\delta>\tfrac12$.
    \item {\bf $b < 1$: coherent impact carries an $O(1)$ share of the variance}. In this case the prefactor
$Y$ carries a power of the metaorder count $M$: it is \emph{intensive} ---
independent of how many metaorders make up the day --- for only one value of $\delta=\tfrac12$, for which $M^{\delta-1/2}=1$ and $\langle x^{2\delta}\rangle=1$ is fixed by conservation, needing no assumption on the size distribution. The closure then collapses to the distribution-free relation
\begin{equation}
\boxed{\;Y^2=1 - b^2.}
\label{eq:Y}
\end{equation}
This is a \emph{partition of unity}: at $\delta=\tfrac12$ the budget reads literally $\sigma^2=Y^2\sigma^2+b^2\sigma^2$, so $Y^2$ is the fraction of daily variance carried by coherent metaorder impact and $b^2$ the complementary fraction carried by idiosyncratic trade-level noise, or by exogenous shocks. That $Y=O(1)$ is now automatic: it is a variance share, necessarily in $[0,1]$, not a quantity that could scale with $M$. For any
other exponent the prefactor scales as $M^{\delta-1/2}$ and cannot be universal
across assets whose metaorder counts differ by orders of magnitude: a linear law
($\delta=1$) would require $Y\propto\sqrt M$, a flat law ($\delta\to0$), $Y\propto
M^{-1/2}$. The observed $O(1)$, roughly asset-independent $Y\approx\tfrac12$ thus
selects the square root --- for reasons of extensive consistency alone.\footnote{Read
conservatively, this argument robustly excludes exponents \emph{far} from
$\tfrac12$; close to $\tfrac12$ the power of $M$ is weak, and pinning the equality
exactly would require the mild extra assumption that $\langle
x^{2\delta}\rangle$ does not drift with $M$.}
\end{itemize}

\paragraph{The mean-equals-standard-deviation identification.} The argument
also settles, rather than assumes, the step we flagged as Bonart's crucial one in
Sec.~\ref{sec:bonart-sqrt}. Setting $\delta=\tfrac12$ in \eqref{eq:impact} and using
$V=N$, $|Q_k|=n_k$, the mean impact of metaorder $k$ is $\varepsilon_k Y s\sqrt{n_k}$
while the standard deviation of its noise term is $b\,s\sqrt{n_k}$: their ratio is\footnote{In fact, this ratio was studied empirically in \cite{bucci_impact_vol2019} (with a misprint,  see the discussion in \cite{maitrier2025subtle}, section A.2) with the result that $Y/b \approx 1/3$. Together with the relation $Y^2 = 1 - b^2$ this leads to $Y \approx 0.32$. Recall again that $Y$ here contains the decay of impact until the end of day, and is therefore expected to be smaller than the commonly reported value $Y \approx 0.5$.}
\begin{equation}
\frac{\text{mean impact}}{\text{std of fluctuations}}=\frac{Y}{b}=O(1),
\qquad\text{independent of } n_k.
\end{equation}
The identification of mean impact with its standard deviation --- postulated in the
diffusivity route --- is here a \emph{consequence} of imposing $\delta=\tfrac12$
together with closure. The volume route thus derives what the whitening route must
assume. Note that when $b \to 1$, $Y \to 0$ and the ratio tends to zero, i.e. in that case, mean impact and standard deviations are not proportional.  

\paragraph{What remains open.} The above reasoning seems to us the correct way to frame Bonart's argument: {\bf if} the impact of metaorders is assumed to
carry an $O(1)$ share of the daily variance, {\bf and if} one imposes
that this share (and hence $Y$) does not scale with the number of contemporaneously
executed metaorders, {\bf then} the SRIL is the only possibility compatible with
volume conservation. What we still lack is a microscopic reason for the premise itself --- for why coherent metaorder impact should retain an $O(1)$,
$M$-independent share of daily variance rather than diluting to the
variance-subdominant $b\to1$ world. But conversely, the fact that the SRIL indeed holds empirically shows that metaorders contribute to $O(1)$ of the volatility, as argued from a different angle in \cite{maitrier2025subtle}.

\section{Two further remarks}
\label{sec:remarks}

We close with two brief remarks placing the above within
the wider scope of the diffusivity constraint. They confirm that
diffusivity alone selects no particular impact law.

\paragraph{A permanent-impact toy-model.} Consider a stationary flow of metaorders
starting at uncorrelated times with uncorrelated signs, each impacting the price
\emph{once}, permanently, at its first child, with no further impact. The
impact-increment series is a train of uncorrelated impulses, cumulative impact is a random walk; returns are diffusive
and unpredictable.  Yet the metaorder impact is permanent, front-loaded, and \emph{independent
of $Q$}: no concave build-up, no $\sqrt Q$. 

A diffusive, information-neutral model can therefore lie arbitrarily far from the SRIL; whether impact scales with $Q$ at
all is set by the amplitude law, not by diffusivity.

\paragraph{The Kyle model.} At the opposite pole, an informed trader's flow is
correlated with fundamental value, the market maker prices competitively, and the
outcome is a diffusive price with impact that is \emph{linear and permanent} \cite{kyle1985continuous}. 

Together with the toy-model above, it shows that diffusivity alone selects neither the flat, the linear, nor the square-root law. Additional, non trivial assumptions must be invoked to pin down the SRIL from diffusivity arguments. 
\section{Conclusion}
\label{sec:conclusion}
Bonart's proposal turns the oldest fact about prices --- their diffusivity --- into a
constraint on market impact, and shows, elegantly, that white impact increments
accumulate to a $\varphi$-independent square root, free of the $\sqrt\varphi$ wart
that afflicts every fixed-kernel model.

What it secures is the additive variance law, under the information-neutrality of
order flow. What it assumes is the step from that law to the square root of a
metaorder --- and that step hides an asymmetry worth stating plainly, for it is the
root of everything else. Diffusivity is a statement about the \emph{variance} of
returns: an additive, stationary quantity, blind to when any particular metaorder
began. The SRIL is a statement about the \emph{mean} impact, which grows as the
square root of the volume executed \emph{since the start of the order} --- an
age-dependent object. One cannot pass from a stationary, clockless variance to an
age-dependent mean without an origin of time; and that origin, the metaorder's
start, is exactly what an anonymous, aggregate tape does not contain. The
Lillo--Mike--Farmer model makes the gap concrete: it satisfies every premise yet,
lacking any such clock, prices an isolated metaorder as the under-concave
$\varphi^{(1-\gamma)/2}Q^{(1+\gamma)/2}$ rather than $\sqrt Q$.

We take Bonart at his
word that his object is not the isolated order but the stationary \emph{stream} of a
participant's metaorders, which the market has learned to whiten --- but that is
indirectly a way of supplying the missing clock. Our multi-agent propagator makes the required ingredient precise, and shows how
strong it is. Whitening each participant's own flow yields the clean law if and
only if its perceived persistence is \emph{linear}, $C^\ast\propto\varphi$ --- and
that linearity is exactly what stream attribution buys: the one missing power of
$\varphi$, relative to the anonymously estimable $\varphi^2$, is the price of
assigning each trade to a coherent, persistent stream and so furnishing it with a
start. The clean SRIL thus requires the market to unmix the aggregate flow into
per-trader streams and whiten each on its own --- a statistical source-separation
problem, not a recovery of identities --- a formidable task that we doubt markets are able to achieve, in view of the so many other persistent anomalies that they are not able to get rid of (see e.g. the recent discussion in \cite{kurth2026trend}). 

The stream hypothesis also makes a sharp, falsifiable prediction --- an isolated
metaorder, well separated in time from its predecessors, should \emph{not} obey the
SRIL --- a claim that is at odds with our available CFM data, which we are revisiting because of the importance of that claim.

To summarize our message: what is in doubt is that whitening
\emph{produces} the square-root law. The
volume-conservation argument of Sec.~\ref{sec:volume} points the same way: among all
exponents, $\delta=\tfrac12$ is the unique one for which the daily variance budget
closes independently of the metaorder-size distribution and of the number of
simultaneous metaorders --- pinned by a conservation law, with no reference to
whitening or detection. The robustness of the square root, we suspect, is a statement
about \emph{volumes that must balance}, not \emph{correlations that must vanish}.
Its natural home is the latent-order-book picture, in which impact is the
displacement required to walk a metaorder's volume through a book of resting
liquidity \cite{mastromatteo2014, donier2015fullyconsistentminimalmodel, donier2016walras, bouchaud2018trades}. The LLOB produces the clean square root natively only at high
participation, whereas the SRIL is measured at low; reconciling the two regimes is
unfinished business, which we take up in a companion paper.

\subsection*{Acknowledgements} We thank Julius Bonart for clarifying email exchanges and comments, which we have tried to take into account in earnest, at least as far as we could agree with them. 

\printbibliography

\end{document}